**Missing Data Imputation and Corrected Statistics for Large-Scale Behavioral Databases**

Pierre COURRIEU and Arnaud REY

Centre National de la Recherche Scientifique – Université de Provence

Running head: Missing data

Corresponding author:

Pierre Courrieu,

Laboratoire de Psychologie Cognitive, UMR CNRS 6146,

Université de Provence, Centre Saint Charles,

Bat. 9, Case D,

3 Place Victor Hugo,

13331 Marseille cedex 3,

France.

E-mail: pierre.courrieu@univ-provence.fr

**Missing Data Imputation and Corrected Statistics for Large-Scale Behavioral Databases**

**Abstract.** This paper presents a new methodology to solve problems resulting from missing data in large-scale item performance behavioral databases. Useful statistics corrected for missing data are described, and a new method of imputation for missing data is proposed. This methodology is applied to the DLP database recently published by Keuleers et al. (2010), which allows us to conclude that this database fulfills the conditions of use of the method recently proposed by Courrieu et al. (2011) to test item performance models. Two application programs in Matlab code are provided for the imputation of missing data in databases, and for the computation of corrected statistics to test models.

## 1. Introduction

An increasing number of large-scale item performance behavioral databases have been published recently, making available a large amount of shared data for building virtual experiments, testing hypotheses and models. In particular, three large-scale databases providing response times and accuracy data for thousands of words, in standard visual word recognition tasks (lexical decision, or naming), are now available for three different languages: the English Lexicon Project (ELP: Balota, Yap, Cortese, Hutchison, Kessler, Loftis, Neely, Nelson, Simpson, & Treiman, 2007), the French Lexicon Project (FLP: Ferrand, New, Brysbaert, Keuleers, Bonin, Méot, Augustinova, & Pallier, 2010), and the Dutch Lexicon Project (DLP: Keuleers, Diependaele, & Brysbaert, 2010).

Examining response time raw data in these databases, one can observe that the amount of missing data is quite important, as a result of response errors, technical failures, or outliers. For instance, there are about 16% missing lexical decision RTs in DLP, which are easily countable since the raw data are simply available and DLP used a complete experimental design where all participants (39) responded to all test words (14089), contrarily to ELP (40481 words) and FLP (39840 words), where each of the numerous participants responded only to a subset of the whole set of test words. Nevertheless, based on the published percents of accuracy, one can estimate that the percent of missing lexical decision RTs is about 16% for ELP, and about 9% for FLP. This is not visible for final users because they usually do not use the raw data. In fact, the databases provide an average RT over all non-missing RTs for each test word, so, missing data are not actually a problem at this stage.

However, in order to test models that predict RT performance at the item level (e.g. Perry, Ziegler, & Zorzi, 2010; Yap & Balota, 2009), it is necessary to know the reproducible proportion of item related variance that is available in the data. It has recently been shown that this proportion is given by a particular intraclass correlation coefficient (ICC), which is the so-called "ICC(C, k), Cases 2 and 2A" coefficient, according to the nomenclature of McGraw and Wong (1996), computed on the raw data table. This method is valid provided that the considered experimental measure fulfills an additive decomposition model that is very commonly assumed (Courrieu, Brand-d'Abrescia, Peereman, Spieler & Rey, 2011; Rey, Courrieu, Schmidt-Weigand & Jacobs, 2009). Courrieu et al. (2011) proposed an efficient test, named ECVT (for "Expected Correlation Validity Test"), to determine the suitability of the ICC approach for any given database. However, this test is based on a Monte-Carlo permutation resampling method, which is sensitive to missing data, and that does not correctly work when the proportion of missing data is large (say, more than 5%). Moreover, it can be shown that missing data act in fact as a parasitic source of noise that harms the data consistency, and, consequently, lowers the reproducible proportion of item-related variance as measured by the ICC. The ICC itself and its confidence intervals can always be computed just using a standard analysis of variance (ANOVA) on any raw data table, and Rey and Courrieu (2010) recently published these statistics for the DLP database. However, without the ECVT test, one cannot be sure that the ICC accurately measures the reproducible proportion of item-related variance.

As we noted above, DLP used a complete experimental design, providing a 14089 items by 39 participants data table on which one can easily compute the ICC (just removing 33 items without valid RT data). However, this is not the case for ELP and FLP databases, where each test word was presented only to a subset of all participants. A possible solution to this problem is to build "virtual participants" by mixing the data provided by several real participants in such a way that each virtual participant has an observation for each item. Using raw data to do this could have catastrophic consequences if the mixed data were provided by real participants having different response characteristics, such as, for instance, their "cognitive speed" (Faust, Balota, Spieler, & Ferraro, 1999). With regard to response latencies, Faust et al. (1999) showed that the inter-individual variability is well modeled by linear relationships. So, a simple way of removing this linear component of the inter-individual variability from the data is to transform the raw data into Z-scores, that is, for each real participant, all his/her valid data are centered by subtracting their average, and divided by their standard deviation. This transformation is near unbiased provided that one assigns with each real participant a large random sample of items. Then Z-scores can be mixed to build virtual participants. So, using Z-scores probably represents a general solution for large-scale databases. However, even using (possibly mixed) Z-scores, there is still a problem with missing data because, in general, the number of valid data is not the same for all items. For instance, in DLP response time database, after removing 33 items with zero valid data, it remains 58 items with only 1 valid data each, 59 items with 2 valid data each, and so on, up to 3219 items with 39 valid data each. This function is plotted in Figure 1.

Statisticians widely studied the problem of missing data, which is often present in large-scale studies. This led them to develop imputation methods that allow replacing missing data with suitable estimates, avoiding introduce statistical biases, as much as possible. Many imputation methods rely on regression-like techniques that are not suitable to the type of data considered here. However, there is at least one imputation method that seems appropriate for our problem. This is the so-called "Adjusted Random Imputation" method (Chen, Rao, & Sitter, 2000), which is very simple and near unbiased, and which, in addition, preserves the original data mean values, thus avoiding modifications of the data usually provided to final users of large-scale databases.

In the next section, we describe the Adjusted Random Imputation (ARI) method of Chen et al. (2000), and we rapidly examine its drawbacks for an application to the type of database considered in this paper. In Section 3, we show that the ICC of data tables with missing data is biased, and we describe a suitable estimate of the "true" ICC of any Z-score type data table with missing data. In Section 4, we describe a new missing data imputation method, called "Column and Row Adjusted Random Imputation" (CRARI), which is a suitable extension of the ARI method that allows adjusting the data ICC. A Matlab program implementing the CRARI method is listed in Appendix A, together with an example of application to real data. In Section 5, we demonstrate, using artificial data, that the ECVT test (Courrieu et al., 2011) provides the same results on data tables without missing data than on similar data tables where a substantial proportion of data has been removed and imputed by the CRARI method. In Section 6, we apply the CRARI imputation method to the DLP database, which then allows us to apply the ECVT test and to conclude positively about the relevance of the ICC approach for this database. In Section 7, we observe that missing data not only degrade the ICC, but they also degrade the average values used as item performance measures. So we propose corrected correlation statistics suitable to solve this problem when one tests models and predictors on the considered data. A Matlab program computing these statistics is listed in Appendix B, together with an example of application to real data. Finally, we conclude in Section 8.

<u>Figure 1</u>

## 2. The Adjusted Random Imputation method

The Adjusted Random Imputation method (ARI: Chen et al., 2000) allows replacing missing data, in a data table, with random estimates in such a way that the average of each row (item) remains the same as in the raw data table, and the expected variance in each row is equal to the observed variance of valid raw data in the same row. Given that the empirical mean and variance are unbiased consistent estimators of the parent parameters, one can hope that the imputation is at least approximately unbiased for the mean and variance. In addition, the method is very simple: for each item, one first replaces each missing data with a valid data randomly selected (with replacement) among all valid data of the considered item. Then one subtracts the average of the replaced data from these data, and one adds to them the average of the valid data, which provides to the imputation the properties above mentioned. See Table 1 for a concrete example.

Table 1

Unfortunately, this elegant method has also some drawbacks that we must take into account. First, in cases where there is only one valid data for a given item, the variance of the data under imputation is zero because only one value is used. Since it is not rare that only one valid data is available for an item in large-scale databases (e.g. this occurs for 58 items in DLP), there is here a possible variance bias. However, the variance in rows, by itself, is not the most important characteristic for the use of item databases. What is actually important, in this case, is the consistency of the data as measured by the ICC, so what we should obtain is an approximately unbiased ICC estimate under missing data imputation.

In order to examine the performance of the ARI method from this point of view, we built artificial data tables, of Z-score type, with known ICCs. Then we gradually degraded these data tables by increasing the proportion of (randomly) missing data. In each case, the Z-scores were recomputed on the degraded data, and the ICC of the obtained data table was computed. Then the ARI imputation method was applied to this table, and the data ICC under imputation was computed. Figure 2 shows a typical example of such an experiment, using 1400-by-80 tables of Z-scores. As one can see in Figure 2, the ICC of the raw data table decreases, and substantially underestimates the exact ICC, as the proportion of missing data increases. On the other hand, the ICC of the data table under imputation by the ARI method increases, and substantially overestimates the exact ICC, as the proportion of missing data increases. Thus, nor the raw data ICC, nor the ICC of data imputed by the ARI method provides a reliable estimate of the exact ICC (i.e., the ICC of the original data table before degradations). In addition, Figure 2 shows a quite accurate estimate of the exact ICC that will be described in the next section.

Figure 2

In summary, the ARI method has the advantage of imputing missing data in a way that preserves the original item means. Unfortunately, this imputation method does not preserve the data consistency as measured by the ICC statistic, and it introduces consistency biases that increase as the proportion of missing data increases. So, we must find another approach.

## 3. Intraclass correlation coefficient corrected for missing data

In this section, we define a suitable estimate of the exact ICC, using the ICC observed on the raw data and the proportion of missing data in the data table. Before doing this, we rapidly recall the data population model that leads to the ICC approach of the reproducible proportion of item related variance (Courrieu et al., 2011; Rey et al., 2009).

### 3.1. Data population model

Let $I$ be a population of items, let $P$ be a population of participants, and let $X$ be a behavioral measure (e.g. response time) on the space $I \times P$, probabilized by a distribution representing, say, the probability for each pair (item, participant) to be selected in an experiment. One assumes that $X$ conforms to the usual additive decomposition model:

$$X = \mu + \alpha + \beta + \varepsilon, \tag{1}$$

where $\mu$ is the mean value of $X$ on $I \times P$, and $\alpha$, $\beta$, and $\varepsilon$ are three independent random variables of mean zero, and of variance $\sigma_\alpha^2$, $\sigma_\beta^2$, and $\sigma_\varepsilon^2$, respectively. The variable $\alpha$ is the participant effect, and it takes a constant value for each given participant. The variable $\beta$ is the item effect, and it takes a constant value for each given item. The variable $\varepsilon$ is considered as a random noise, however, it can as well result from the combination of an item-participant interaction and of a true random noise. The variable $\beta$, whose values characterize the items, is the variable of interest in this study.

One can derive from $X$ another measure, denoted $X^{(n)}$, that is the arithmetic mean of $X$ over $n$ randomly selected distinct participants (thus $X^{(1)} = X$), then one obtains from (1) the following decomposition:

$$X^{(n)} = \mu + \alpha^{(n)} + \beta + \varepsilon^{(n)}, \tag{2}$$

where the random variables $\alpha^{(n)}$, $\beta$, and $\varepsilon^{(n)}$ are always independent with means zero, but their variances are now $\sigma_\alpha^2 / n$, $\sigma_\beta^2$, and $\sigma_\varepsilon^2 / n$, respectively.

Consider now the bivariate distribution of pairs $(x, y)$, where $x$ and $y$ are independent realizations of $X^{(n)}$. Then the population correlation between $x$ and $y$, varying the items, is given by:

$$\rho(x, y) = \frac{\sigma_\beta^2}{\sigma_\beta^2 + \sigma_\varepsilon^2 / n}. \tag{3}$$

One can recognize in (3) the expression of a well-know intraclass correlation coefficient (ICC), that is the "ICC(C, k), Cases 2 and 2A" coefficient, according to the nomenclature of McGraw and Wong (1996). The expression (3) itself shows that the ICC is the proportion of systematic variance ($\sigma_\beta^2$) in the total item related variance ($\sigma_\beta^2 + \sigma_\varepsilon^2 / n$), which also follows from Equations (12-13) in Courrieu et al. (2011).

### 3.2. Corrected ICC

We consider now random variables similar to those defined in (2), but where a proportion $p$ of values are randomly missing. Then the mean number of averaged values is no longer $n$, but:

$$n' = (1 - p)n \tag{4}$$

Denoting $\rho_p$ the ICC for variables with a proportion $p$ of randomly missing values, and using (3) and (4), one has approximately:

$$\rho_p \approx \frac{\sigma_\beta^2}{\sigma_\beta^2 + \sigma_\varepsilon^2 / n'} \tag{5}$$

So we can write:

$$1/\rho_0 - 1/\rho_p \approx \frac{\sigma_\varepsilon^2}{\sigma_\beta^2} \left(1/n - 1/n'\right) = \frac{n'(1 - \rho_p)}{\rho_p} \left(1/n - 1/n'\right),$$

where $\rho_0$ is the exact ICC (without missing values). It follows from the above relation that:

$$\rho_0 \approx \left( \frac{n'(1 - \rho_p)}{\rho_p} \left(1/n - 1/n'\right) + \frac{1}{\rho_p} \right)^{-1} = \frac{\rho_p}{1 - p(1 - \rho_p)} \geq \rho_p, \tag{6}$$

where the last inequality is strict if $p > 0$ and $\rho_p < 1$, which means that $\rho_p$ underestimates $\rho_0$. From (6), one can also define an estimate of $\rho_0$ as:

$$\rho_{cor} = \frac{\rho_p}{1 - p(1 - \rho_p)}, \tag{7}$$

where the subscript "$cor$" means "corrected" (for missing data). In practice, $\rho_p$ is estimated by the ICC of the data table, computed by a standard ANOVA, and $p$ is the observed proportion of missing data in this table.

One does not obtain $\rho_{cor} = \rho_0$, in general, because randomly missing values induce contaminations between the different sources of variation, in such a way that (5) is a rough approximation if $\sigma_\alpha^2$ is not negligible. However, if $\sigma_\alpha^2$ is small with respect to both $\sigma_\beta^2$ and $\sigma_\varepsilon^2$, then $\rho_{cor}$ appears to be a suitable estimate of $\rho_0$. Fortunately, $\sigma_\alpha^2$ is just the variance corresponding to the column (participant) effect in a data table, and thus this variance is removed when one uses Z-scores as behavioral measures. In order to test the suitability of $\rho_{cor}$ as an estimate of $\rho_0$, we built artificial data tables gradually degraded as in Section 2, but instead of considering only Z-scores, we used four different versions of each table: a version with non-centered columns (and a substantial $\sigma_\alpha^2$), a version with centered columns (thus $\sigma_\alpha^2 = 0$), a Z-scores version, and a mixed Z-scores version where the Z-data in each row where randomly mixed. One can see in Figure 3 that, when the columns of the data tables are not centered, the estimate ($\rho_{cor}$) has a positive bias that increases as the proportion of missing data increases. However, when the columns are centered, as well as for Z-scores and mixed Z-scores, the estimate is visibly reliable since it randomly oscillates in a close neighborhood of the exact value for all proportions of missing data (up to 30% in these examples, which is much greater than percentages of missing data commonly observed in real databases). Note that all plots in Figure 3 are identical, except the first one. In particular, randomly mixing Z-scores in each row of a data table preserves the ICC statistics with respect to the non-mixed case. As a practical rule for the column effect, we observed that $\rho_{cor}$ is a reliable estimate of $\rho_0$ if there is no more than 5% missing data, or if the estimated column effect ($s_\alpha^2$) is not greater than both the row effect ($s_\beta^2$) and the row-by-column interaction ($s_\varepsilon^2$). This rule is implemented in the application programs listed in Appendix A and Appendix B.

<u>Figure 3</u>

In summary, we defined a simple and easy to compute statistic, denoted $\rho_{cor}$, which is an ICC corrected for missing data, and which is an estimate of the ICC that would be obtained if no data were missing. This estimate is reliable provided that the column effect in the considered data table is small or removed.

### 3.3. Building virtual participants from datasets with incomplete designs

As we observed in Section 3.2, randomly mixing Z-scores in each row of a data table preserves the ICC statistics with respect to the non-mixed case. This suggests a possible strategy for structuring datasets with incomplete designs (e.g. ELP, or FLP), in order to make possible the computation of their ICC statistics. First, one must transform the raw data into Z-scores in order to remove the linear component of the individual variability. Let $n$ be the observed maximum number of valid data per item, then one can build a rectangular data table with $m$ items (rows) and $n$ "virtual participants" (columns), and one randomly assigns each valid Z-score to a column, in the appropriate row. In general, this leaves "holes" in the table because there are items with less than $n$ valid data. These holes can be treated as ordinary missing data, and the ICC statistics can be computed on the table as if we had $n$ real participants. This approach will not be developed more in this article, and it is just mentioned as an indication for further investigations.

## 4. Column and Row Adjusted Random Imputation method

In this section, we define a new imputation algorithm, called " Column and Row Adjusted Random Imputation " (CRARI), that allows replacing missing data by imputed values in such a way that the resulting item means are the same as those of the initial data table, and the ICC of the data table under imputation can be set to any desired value in a wide range of possible values, including the ICC of the initial data table (approximately $\rho_p$), and the estimate $\rho_{cor}$ of $\rho_0$ defined by (7), when appropriate. In particular, this last option is always suitable for tables of Z-scores, as stated in Section 3. The CRARI algorithm is an extension of the ARI algorithm of Chen et al. (2000), however, it avoids the biases of the ARI method observed in Section 2 for items-by-participants data tables of Z-scores.

### 4.1. Principle of the method

First, we note that the CRARI algorithm, just as the ARI algorithm, is deterministic whenever there is no more than one missing data per item in the raw data table. In this case, an imputed value, for a given item, is necessarily equal to the average of all valid observations for that item, in order to preserve the item mean. Thus, in the following, we assume that at least one item has more than one missing data in the raw data table, which is usually the case in large-scale databases.

The first problem to be solved is the case of items with only one valid data, which leads to impute values with zero variance when using the ARI method. So, the first step of the CRARI method consists in applying the ARI algorithm to each column (instead of each row) of the data table, which results in a provisional imputed data table where all column means are unchanged, but each row (item) contains values with a non-zero expected variance, even if only one valid data is available for that item. At this stage, the item means are not adjusted, that is, they are (probably) different from those of the initial data table. The second step of the CRARI method operates on each row of the provisional data table, where all imputed values (if any) must be centered by subtracting their average. Then one multiplies them by a positive coefficient $c$ whose computation will be described below, and one add to the imputed values the average of the valid data available for the considered item. As a result, whatever be the coefficient $c$, the item means are now equal to those of the initial data table, leaving also the general mean of the table unchanged. However, the resulting column means can be somewhat different from those of the initial table, despite the fact that their expected value is unchanged. A suitable choice of the coefficient $c$ allows adjusting the ICC of the data table under imputation, while this ICC can be arbitrarily chosen in a wide range of values. Two non-arbitrary ICC values are of special interest: the ICC of the initial data table (hereafter called "low ICC"), which can always be computed by an ANOVA, and the estimate $\rho_{cor}$ that can be computed using the low ICC as $\rho_p$ in (7). Whatever be the target ICC, one can reach it thanks to the following mechanism. Provided that we consider several distinct imputed values for an item, their (non-zero) variance is multiplied by $c^2$, and thus, globally, the variance of the data associated to that item increases as $c$ increases, or it decreases as $c$ decreases. Part of this variance affects the column effect, which is not of interest, but the remaining part of the adjustable variance affects the row-by-column interaction effect, which is just what we need to control the ICC, given that the row effect remains unchanged due to the fact that the item means are fixed. As a result, the ratio $q = \sigma_\beta^2 / \sigma_\varepsilon^2$ decreases as $c$ increases, and conversely $q$ increases as $c$ decreases. Given that the ICC is a monotonic increasing function of the $q$ ratio, we have a simple mean to control the ICC value of the data table under imputation. So, one can use a simple dichotomic search procedure to compute the $c$ value providing the target ICC value, which is the method implemented in the Matlab program CRARI listed in Appendix A. This program can be used directly, or as an implementation model. For readers not familiar with Matlab code, we summarize hereafter the method in pseudo-code.

### *4.2. The CRARI algorithm*

Given:

- a data table of ($m$ items)-by-($n$ participants), with missing data,

- and a *Target ICC* value ( $\rho_p$ computed by ANOVA, or $\rho_{cor}$ computed from $\rho_p$ by (7))

**If** (no more than one missing data per item) **then**                    {deterministic case}

      **For** *item* $\leftarrow$ 1 **to** $m$

            Replace the missing data, if any, by the average of all valid data in the row.

      **End**

      The output is the resulting table

**Else**                                                              {random imputation}

      **For** *participant* $\leftarrow$ 1 **to** $n$

            **For** each missing data in the column

                  randomly replace the missing data by a valid data from the column.

            **End**

            Compute the average of the replaced data in the column.

            Subtract this average from the replaced data in the column.

            Compute the average of all valid data in the column.

            Add this average to the replaced data in the column.

      **End**

      **For** *item* $\leftarrow$ 1 **to** $m$

            Compute the average of all replaced data (if any) in the row.

            Subtract this average from the replaced data in the row.

**End**

$c_{min} \leftarrow 0$, $c_{max} \leftarrow$ maximal $c$ (e.g. 10), *tolerance*←small positive number (e.g. $10^{-4}$)

**Repeat**

$c \leftarrow (c_{min} + c_{max}) / 2$

Compute a provisional table where:

all valid data remain unchanged,

and all replaced data are multiplied by $c$.

Compute the *ICC* of the provisional table by a standard ANOVA.

**If** (*ICC > Target ICC*) **then**

$c_{min} \leftarrow c$

**Else**

$c_{max} \leftarrow c$

**End**

**Until** ($c_{max}$ - $c_{min}$) < *tolerance*

The output is the last computed provisional table.

**End**

### 4.3. Behavior of the imputation method

In order to test the behavior of the new method, we replicated the experiments on artificial data of Section 2, but using the CRARI method, instead of the ARI method, for imputation. Figure 4 shows an example of experiment where the target ICC was the estimate $\rho_{cor}$. As one can see, the CRARI algorithm reached the target ICC in all cases, providing a close approximation of the exact ICC. Now, if one takes the low ICC as target, then one obtains an "imputed" curve undistinguishable from the "missing" curve of Figure 4. Thus, clearly, the CRARI algorithm is efficient for imputation of data with a prescribed ICC.

Figure 4

In summary, the CRARI method, just as the ARI method, has the advantage of imputing missing data in a way that preserves the original item means. In addition, the CRARI imputation method allows us to preserve the data consistency as measured by the ICC statistics, while one can choose the "low ICC" observed with missing data, or the ICC corrected for missing data ($\rho_{cor}$), as the ICC of the data table under imputation.

## 5. The ECVT test on data tables with imputation of missing data

A major interest of having an efficient imputation method is to allow the use of permutation resampling techniques that do not work on data tables with too many missing data. This is the case of the ECVT test (Courrieu et al. 2011), which was shown to efficiently test the compatibility of any data table with the additive decomposition model (1), and consequently, the relevance of the ICC approach to determine the proportion of item related variance that models should account for in the considered data set. This test cannot work, for instance, on the DLP database (Keuleers et al., 2010) because missing data (16%) prevent from obtaining pairs of complete vectors of item means necessary to the permutation resampling procedure of the ECVT test. This was, in fact, our first motivation to develop the CRARI method, however, before using it in conjunction with the ECVT test, we must be sure that the imputation does not induce biases that could lead the ECVT test to provide wrong conclusions.

### 5.1. Building artificial data

A suitable way of examining this problem consists in building complete artificial data tables that fulfill or do not fulfill model (1), and degraded versions of the same data tables, including a substantial proportion of missing data, to which one applies the CRARI algorithm in order to impute the missing data. Then, applying the ECVT test to both data tables, one must obtain the same conclusions for the tables with imputed data as for the original tables. It is easy to build artificial data that fulfill model (1), however, building data that do not fulfill model (1), we must take care that the discrepancy of the data from model (1) cannot be removed by the Z-score transformation, since the CRARI method is mainly devoted to work on Z-scores. One can obtain suitable data using the following generating process:

$$x_{ij} = \mu + sign(\beta_i) \times |\beta_i|^{\alpha_j} + \varepsilon_{ij}, \tag{8}$$

where $i$ is the index for the row (item), and $j$ is the index for the column (participant). The variables $\mu$, $\beta$, and $\varepsilon$ are defined as in model (1), however, the participant effect ($\alpha$) is not additive, nor multiplicative since it would be removed by the Z-score transformation. The participant effect is a random variable generated by the following process:

$$\alpha_j = 1 - s \times \log(1 - u_j), \qquad (9)$$

where $u_j$ is uniformly randomly sampled in the interval [0,1), and $s$ is a non-negative parameter. Observe that if $s = 0$ then the participant effect is always equal to 1, and the generated data fulfill model (1). However, the discrepancy of the generated data from model (1) increases as $s$ increases, as a result of using powers different from 1 in (8). The cumulative probability function of $\alpha$ in this model is given by:

$$P(\alpha) = 1 - \exp(-(\alpha - 1)/s), \quad \alpha \geq 1. \qquad (10)$$

### 5.2. Randomly distributed missing data

We randomly generated data tables of size 1400×80 using model (8), and varying the parameter $s$ of (9). For each table, we built a degraded version of the table with 16% randomly missing data, and we transformed the data of both tables in Z-scores. Then the CRARI algorithm was applied to the degraded tables with $\rho_{cor}$ as the target ICC, resulting in data tables with 16% imputed data. Finally, we applied the ECVT test to both data tables (for a detailed description of this test, see Courrieu et al. 2011). Figure 5 shows an example of ECVT results for data tables generated with $s = 0$, thus compatible with model (1), and for data tables generated with $s = 2$, thus incompatible with model (1). As one can see, in all cases, the ECVT test provided the correct conclusion, which was always the same for the original data table and for the corresponding imputed data table. Lowering $s$ to 1, one still obtains correct detections of the discrepancy from model (1) for both original and imputed data tables, although the discrepancy is no longer visible on the test graphs. As previously noted by Courrieu et al. (2011), the ECVT test is more sensitive than the eye. We conclude from this study that the ECVT test provides similar conclusions for original data tables and for data tables with a substantial proportion of data imputed by the CRARI algorithm, when the missing data are randomly distributed.

Figure 5

### 5.3. Non-randomly distributed missing data

Randomly distributed missing data can occur in certain databases, for instance when missing data mainly result from technical failures. However, in experimental paradigms such as lexical decision or speeded naming, missing data frequently result from response errors, and these errors are not random since they typically concentrate on items with low frequency of use and special difficulties (Balota, Cortese, Sergent-Marshall, Spieler, & Yap, 2004), which are also the items with the greatest response times (when not missing). Since the DLP database includes lexical decision times, it appeared necessary to verify that the ECVT test, combined with the CRARI imputation method, provides suitable conclusions on data tables with a distribution of missing data similar to that of DLP. In order to do this, we first reordered the rows (items) of the DLP database by increasing order of the item mean RTs transformed in Z-scores. Then the exact locations of all missing data in the resulting table were marked. Using (8), we generated complete artificial data tables of the same size as the DLP data table (14056 by 39), varying the $s$ parameter of (9) from 0 to 6, and using a $q$ ratio approximately equal to that of the DLP Z-scores (for $s = 0$). The rows of each table were reordered by increasing order of their means, and a degraded version of each table was obtained by removing the data at the same locations as the missing data in the reordered DLP data table. Then both tables were transformed in Z-scores, and the CRARI algorithm was applied for imputation of the missing data to the degraded tables, with $\rho_{cor}$ as the target ICC, and the ECVT test was applied to both resulting tables. With such data, it appeared necessary to increase the $s$ parameter of (9) up to 6 in order to obtain detectable discrepancies from model (1). Outcomes of the ECVT test in this experiment are plotted in Figure 6, where one can see that the conclusions of the ECVT test are correct for $s = 0$ (compatibility with model (1)), as well as for $s = 6$ (discrepancy from model (1)), for both the original complete data tables and the corresponding data tables under imputation for missing data. However, one can note in Figure 6 that the standard deviations of the $r$ distributions in case of discrepancy from model (1) are substantially lower for the data table under imputation than for the original data table, which suggests that the imputation is not unbiased in terms of distribution. Nevertheless, the ECVT test seems to provide reliable conclusions, even when combined with the CRARI imputation for missing data.

Figure 6

In summary, we observed that the ECVT test provides the same conclusions for data tables without missing data than for similar data tables where a substantial proportion of data has been replaced with data imputed using the CRARI method. This is true for randomly distributed missing data, as well as for more realistic distributions of missing data that mimic the one of the DLP database.

## 6. Application to the DLP database

We computed the ICC and three of its confidence intervals (95%, 99%, and 99.9%) for four versions of the DLP response time database: the raw data (RTs), the Z-scores, the Z-scores with imputation of missing data (16%) by CRARI with the low ICC, and Z-scores with imputation of missing data by CRARI with the $\rho_{cor}$ ICC. The results are reported in Table 2, where one can see that the statistics, including the confidence intervals, are exactly the same for Z-scores and imputed Z-scores with low ICC. We note also that using Z-scores (imputed or not) improves the reproducible proportion of item related variance with respect to raw data, and this improvement is significant since the confidence intervals do not overlap. Note, however, that this last result does not generalize to every database. Depending on each particular data distribution, Z-scores are sometimes advantageous, and sometimes they are not, as one can verify using simulated data. The imputation of missing data using $\rho_{cor}$ clearly provides the most consistent data table. Moreover, if one directly applies (7) to the confidence limits of the ICC of the DLP Z-scores, without imputation for missing data, one obtains the confidence intervals [0.879, 0.884], [0.878, 0.885], and [0.877, 0.886] for 95%, 99%, and 99.9% confidence, respectively. These confidence intervals are equal, with three decimal digits, to those obtained with imputation for missing data with $\rho_{cor}$ as the target ICC. This provides a rapid way of estimating the confidence intervals of $\rho_{cor}$ without imputation. At this point, the reader probably ask to what ICC the proportion of variance accounted for by a model must be compared. We break the suspense: use the raw data ICC if the model correlation is computed with the average raw data, and use the Z-scores ICC if the model correlation is computed with the average Z-scores. The $\rho_{cor}$ ICC is useless to test models (we will see why in Section 7), except if one uses the $r_{cor}^2$ goodness of fit statistic defined in Section 7.3. However, as demonstrated in Section 5, it is relevant to impute missing data using $\rho_{cor}$ as the target ICC, in order to be able to test properties of the data themselves.

Table 2

This is just what we do now, applying the ECVT test to the DLP Z-scores with CRARI imputation of missing data, using $\rho_{cor}$ as the target ICC. The result of the ECVT test can be shown in Figure 7, where one can see that the theoretical and observed correlation curves are undistinguishable, and the $\chi^2$ test is non-significant, indicating that the DLP data are compatible with model (1).

Figure 7

As a supplemental verification, we applied the ECVT test to the DLP Z-scores with CRARI imputation of missing data, but using the low ICC as the target. The result of the ECVT test can be shown in Figure 8, where one can see that, once again, the theoretical and observed correlation curves are undistinguishable, and the $\chi^2$ test is non-significant, confirming that the DLP data are compatible with model (1). So, the statistics reported in Table 2 can be confidently used to test models with the DLP database.

Figure 8

The validity of the ICC approach had been previously demonstrated, using the ECVT test, for word identification times in English, and for word naming times in English and in French (Courrieu et al., 2011). However, it is the first time that we can conclude with regard to lexical decision times (in Dutch), which is very important since lexical decision is the most widely used experimental paradigm in visual word recognition studies. An alternative approach would be to remove all items presenting a high proportion of missing data, and to apply the ECVT test only on the remaining items, without imputation for missing data. This approach leads to remove a large proportion of items with low frequency of use, which is not desirable because rare words are useful in experiments, where their relative frequency of occurrence is much greater than in the every day life. Moreover, statistics computed only on the most frequent items are not necessarily good estimates for the whole item population.

In summary, we provided the ICC statistics (with confidence intervals) for the lexical decision times of the DLP database and their Z-score transformation. Using the CRARI imputation method allowed us to apply the ECVT test to DLP Z-scores, and to conclude that the ICC suitably measures the reproducible proportion of item related variance that models should try to account for in the Z-score version of the DLP database.

## 7. Compensating item means inaccuracy for model tests

### 7.1. Missing data induce inaccuracy in item means

It is well known that the empirical mean of a random sample of size $n$ is an unbiased, consistent estimator of the parent mean. More precisely, as $n$ increases, the estimation error of the mean rapidly tends to be normally distributed with mean 0, and variance $\left(E(v^2) - E^2(v)\right)/n$, where $v$ is the considered random variable, and $E$ is the usual "expected value" operator. Thus, preserving the empirical item means in imputation methods is certainly a good strategy because empirical means are unbiased estimators of the parent means, however, the accuracy of the estimates clearly depends on the number of empirical data actually averaged for each item, while this number is affected by missing data. As a result, the accuracy of the empirical vector of item means decreases as the proportion of missing data increases in the data table. This is illustrated in Figure 9, were the results of an experiment with artificial data similar to those of Section 2 and Section 4 are plotted, together with the correlation between the vector of item means of the original data table, and the one of degraded data tables with an increasing proportion of missing data (up to 90%). As one can see, the degradation of the correlation of item means ("r(item means)" curve) is roughly parallel to the degradation of the low ICC ("missing" curve). In the same time, the $\rho_{cor}$ ICC ("estimate" curve), and the CRARI imputation ICC ("imputed" curve) remain equal and oscillate in a neighborhood of the exact ICC ("exact" curve), with a moderately increasing variance. This is the reason why the low ICC is usually the good reference to test models, since one uses the (squared) correlation between model predictions and possibly degraded empirical item means, according to Courrieu et al. (2011).

Figure 9

### 7.2. Virtues of the $r^2/ICC$ ratio

However, there is another way of testing predictor type models, and that way is, in a sense, more general than the simple model misfit test proposed in Courrieu et al. (2011). Noting that the usual squared Pearson's correlation coefficient $r^2$ (or $R^2$, for multiple regression models) is an estimate of the proportion of item related variance accounted for by a given model, we deduce that the ratio $r^2/ICC$ estimates the proportion of reproducible item related variance that is accounted for by the model, given that the ICC estimates the reproducible proportion of item related variance actually available in the data. But there is more, as we state now.

The parent ratio of $r^2/ICC$ is $\rho^2(x,B)/\rho(x,y)$, where $B$ is the considered predictive variable (model prediction), $x$ and $y$ are two independent realizations of $X^{(n)}$, as in (3). Remembering that $Var(x) = Var(y) = \sigma_\beta^2 + \sigma_\varepsilon^2/n$ (see Courrieu et al., 2011), one can write:

$$\rho^2(x,B)/\rho(x,y) = \frac{Cov^2(x,B)}{Var(x)Var(B)} \bigg/ \frac{Cov(x,y)}{\left(Var(x)Var(y)\right)^{1/2}} = \frac{Cov^2(x,B)}{Var(B)Cov(x,y)},$$

and given that

$$Cov(x,B) = Cov(\beta,B) + Cov(\varepsilon^{(n)},B), \ Cov(x,y) = \sigma_\beta^2, \text{ and } Var(B) = \sigma_B^2,$$

one obtains

$$\rho^2(x,B)/\rho(x,y) = \frac{\left(Cov(\beta,B) + Cov(\varepsilon^{(n)},B)\right)^2}{\sigma_\beta^2 \sigma_B^2}. \tag{11}$$

If $Cov(\varepsilon^{(n)},B) = 0$ then

$$\rho^2(x,B)/\rho(x,y) = \frac{Cov^2(\beta,B)}{\sigma_\beta^2 \sigma_B^2} = \rho^2(\beta,B), \tag{12}$$

which is just the squared correlation between the predictive variable $B$ and the hidden behavioral variable $\beta$. Note that this quantity is independent of the noise and of the number of participants, which makes it of special interest for various purposes, and in particular for removing the effect of missing data from model goodness of fit statistics. The case of (12) is what normally occurs if the predictive variable $B$ does not over-fit the data.

However, if $Cov(\beta,B)Cov(\varepsilon^{(n)},B) > 0$ then (11) implies that $\rho^2(x,B)/\rho(x,y) > \rho^2(\beta,B)$. This case typically corresponds to an over-fitting, that is, the predictive variable $B$ is correlated with the data noise (with the same sign as its correlation with $\beta$), generally due to the use of too many free parameters in the model to fit the data.

Now, if $Cov(\beta,B)Cov(\varepsilon^{(n)},B) < 0$ then (11) implies that $\rho^2(x,B)/\rho(x,y) < \rho^2(\beta,B)$, however, one can hardly imagine a mechanism generating such a situation (negative over-fitting?), except hazard.

In order to illustrate (12), we considered two behavioral databases previously used in Courrieu et al. (2011), and in Rey, Brand-d'Abrescia, Peereman, Spieler, and Courrieu (2010). These databases are an English word naming RTs table of size 770 items by 94 participants, with 3.61% missing data, and a French word naming RTs table of size 615 items by 100 participants, with 3.94% missing data. For each database, we considered two well-known predictors, that is, the word log-frequency of use, and the word length (number of letters). We used a permutation resampling procedure with various participant group sizes, similar to the procedure used in the ECVT test. However, in addition to the computation of the correlation between item means of two groups, at each resampling step, we also computed the correlation between item means of one group and each of the two predictors. At the end, we obtained, for each participant group size, an estimate of the ICC, and estimates of the correlations between item means and the two predictors. Then, for each group size and for each predictor, we computed the ratio $r^2/ICC$ and we plotted the obtained ratios as functions of the participant group size in Figure 10, for the two databases. We know that the ICC is a monotonic increasing function of the number of participants, however, (12) predicts that $r^2$ increases in the same way as the ICC, in such a way that the ratio $r^2/ICC$ remains constant. This is just what we can observe in Figure 10 for the two databases and the two predictors. Incidentally, we can also observe that the word frequency effect is stronger than the word length effect in English, while the converse is true in French.

Figure 10

### 7.3. Corrected $r^2$ goodness of fit statistic

Here, we exploit the fact that (12) is independent of the noise and of the number of participants. Let $r_p$ be the correlation coefficient of a predictive variable $B$ with a behavioral measure, averaged over $n$ participants with a proportion $p$ of missing values, and let $\rho_p$ denote the corresponding ICC. If there is no over-fitting, then one has after (12):

$$r_p^2 / \rho_p = r^2(\beta, B), \tag{13}$$

and in particular for $p = 0$, one has:

$$r_0^2 / \rho_0 = r^2(\beta, B) \implies r_0^2 = \rho_0 \, r^2(\beta, B). \tag{14}$$

Combining (13) and (14) one obtains:

$$r_0^2 = \rho_0 \, r_p^2 / \rho_p. \tag{15}$$

Now, in cases where $\rho_{cor}$ is a suitable estimate of $\rho_0$ (e.g. Z-score type data), we have:

$$r_0^2 \approx \rho_{cor} \, r_p^2 / \rho_p, \tag{16}$$

that is, multiplying the observed ratio $r^2 / ICC$ by the ICC corrected for missing data ($\rho_{cor}$ defined by (7)), we obtain an estimate of the squared correlation of the predictive variable $B$ with the behavioral measure averaged over $n$ participants without missing data. So, using (16), we define the $r^2$ statistic corrected for missing data as:

$$r_{cor}^2 = \rho_{cor} \, r_p^2 / \rho_p. \tag{17}$$

Note that the reference ICC for the goodness of fit statistic $r_{cor}^2$ is $\rho_{cor}$, whose confidence intervals can be computed on a data table under CRARI imputation with $\rho_{cor}$ as the target ICC, or can be directly approximated using (7) on the ICC confidence limits. The Matlab program ICCR2 listed in Appendix B takes as arguments a data table and a set of predictors, and it provides as outputs the ICC statistics of the data table and the goodness of fit statistics of the predictors ($r^2$, $r^2 / ICC$, and $r_{cor}^2$). An example of use of the program is also provided. The goodness of fit statistics can also be used as complementary information in multiple regression analyses, remembering that $R^2$ is equal to the squared correlation coefficient of the data with the optimal composite predictor computed by multiple regression analysis (Cohen, Cohen, West, & Aiken, 2003).

In order to illustrate the behavior of the $r_{cor}^2$ estimate, we built artificial data tables as in Section 7.1, and an artificial predictor with known $r^2$ goodness of fit. Then we increased the proportion of missing data up to 90%, and we plotted $r_0^2$ ("exact" curve), $r_p^2$ ("observed" curve), and $r_{cor}^2$ ("estimate" curve), as functions of the percentage of missing data, in Figure 11. As one can see, $r_{cor}^2$ behaves as a near unbiased estimator of $r_0^2$, however, the estimator variance increases as the proportion of missing data increases. Nevertheless, the estimator variance remains small enough for practical use with commonly observed percentages of missing data. Thus, we can conclude that $r_{cor}^2$ is a suitable estimate for practical applications, while the bias of the observed $r_p^2$ is visibly substantial and dramatically increasing as the proportion of missing data increases. So, we can be sure that $r_{cor}^2$ is at least a much better estimate than the usual $r^2$ in cases where there are missing data, provided that one uses Z-score type data.

<u>Figure 11</u>

In summary, after observing that missing data degrade the accuracy of the averages used as item performance measures, in the same way as they degrade the data consistency, we proposed two useful goodness-of-fit statistics to test models: the ratio $r^2/ICC$, which is approximately independent of the data accuracy and consistency, and the $r_{cor}^2$ statistic, which is a version of the usual $r^2$ statistic corrected for missing data. The $r_{cor}^2$ statistic can be compared to $\rho_{cor}$ in the same way as the usual $r^2$ statistic can be compared to the low ICC.

## 8. Conclusion

This study is in the continuation of the work recently published by Courrieu et al. (2011), which addressed the problem of the amount of variance that item level performance models should account for in large-scale behavioral databases. In the previous study, it was shown that the reproducible proportion of item related variance in such databases is suitably measured by a particular intraclass correlation coefficient (ICC) computed on the data table, provided that the considered behavioral measure fulfills a usual additive decomposition model. Then a powerful test, named ECVT, was proposed to detect discrepancies of the considered data from this model, and it was shown that commonly used behavioral measures suitably fulfill the model, making the ICC approach relevant. However, the ECVT test is based on a permutation resampling procedure that cannot correctly work when there are too many missing data in the considered data table. Unfortunately, the proportion of missing data is commonly quite large in large-scale databases, due to response errors, outliers, and technical failures. So, in the present study, we addressed the problem of missing data and we proposed ways of overcoming this problem. We noted that the Z-score transformation of raw data (Faust et al., 1999) is a general solution for suitably formatting large-scale behavioral databases, so the present study mainly focused on Z-score type data.

The first idea that one can have to solve the problem of missing data is to replace the missing data with suitable estimates that preserve the essential statistical properties of the whole data set, without introducing biases. We first described the Adjusted Random Imputation (ARI) method of Chen et al. (2000), and using artificial data, we observed some drawbacks of this method for an application to large-scale item level behavioral databases. Then, we showed that the ICC of data tables with missing data is biased, and we described a suitable estimate of the exact ICC of any Z-score type data table with missing data. We described a new missing data imputation method, called "Column and Row Adjusted Random Imputation" (CRARI), which is a suitable extension of the ARI method that allows adjusting the data ICC. A Matlab program implementing the CRARI method is listed in Appendix A, together with an example of application to real data.

Then, we demonstrated, using artificial data, that the ECVT test proposed by Courrieu et al. (2011) provides the same results on data tables without missing data than on similar data tables where a substantial proportion of data has been removed and imputed by the CRARI method. We applied the CRARI imputation method to the DLP database (Keuleers et al., 2010), which then allowed us to apply the ECVT test to this database under imputation of missing data, and to conclude that the ICC approach is relevant for this database. This is the first time that the approach of Courrieu et al. (2011) is validated on lexical decision times, which is an important result since lexical decision is the most widely used experimental paradigm in visual word recognition studies. The ICC statistics of the DLP database were provided for practical use in model testing on these data.

However, we observed that missing data not only degrade the data table ICC, but they also degrade the average values used as item performance measures. So, we proposed corrected correlation statistics suitable to solve this problem when one tests models and predictors on data sets with missing data. A Matlab program computing these statistics is listed in Appendix B, together with an example of application to real data. Note that users having difficulties for using the provided Matlab programs can directly contact the authors (pierre.courrieu@univ-provence.fr).

We conclude that this study provides effective and reasonably robust methods to solve the problem of missing data that always occurs in large-scale studies, and to test item performance models and predictors on the large-scale behavioral databases recently made available to the community of researchers.

**Appendix A.** CRARI program (Matlab 7.5 code) for missing data imputation. Texts at right of "%" are comments.

```
-----------------------------------------------------------------------------------------
function [IX,c,iccX,iccCor,iccIX] = CRARI(X,MissCode,LowICC)
% ----------- Column and Row Adjusted Random Imputation ------------------
%                            Input:
% X: (m x n) data table; MissCode: code for missing data (default: Inf);
% If LowICC=0 (default), then impute using the corrected ICC (iccCor),
%    else impute using the ICC of the data table (iccX).
%                            Output:
% IX: (m x n) imputed data table
% c: c coefficient computed by the CRARI algorithm
% iccX: ICC of X; iccCor: ICC corrected for missing data; iccIX: ICC of IX
%                            Note:
% There must be at least one data per row and column, and if LowICC=0, then
% X columns must have approximately equal means, as with (mixed) Z-scores.
% -------------------------------------------------------------------

if nargin<3, LowICC=0; end, if nargin<2, MissCode=inf; end      % Default

[m,n]=size(X);                                        % ANOVA of X
ti = zeros(m,1); ni = ti; tj = zeros(1,n); nj = tj;
sx2 = 0;
for i = 1:m
    for j = 1:n
        if X(i,j) ~= MissCode
            ti(i,1) = ti(i,1)+X(i,j);
            ni(i,1) = ni(i,1)+1;
            tj(1,j) = tj(1,j)+X(i,j);
            nj(1,j) = nj(1,j)+1;
            sx2 = sx2 + X(i,j)^2;
        end
    end
end
if ~isempty(find(ni==0,1)), error('Error in CRARI: empty row(s)'), end;
if ~isempty(find(nj==0,1)), error('Error in CRARI: empty column(s)'), end;
mitemX = ti./ni; maxmiss=max(n-ni);
N = sum(ni); pmiss=(m*n-N)/(m*n);        % Proportion of missing data in X
t = sum(ti); ss = sx2 - t^2/N;
ssi = sum(ti.^2./ni) - t^2/N;
ssj = sum((tj.^2./nj),2) - t^2/N;
ssij = ss - ssi - ssj;
dfi = m-1; dfj = n-1; dfij = N-1-dfi-dfj;
msi = ssi/dfi; msj=ssj/dfj;
vij = ssij/dfij; vi = max(0,(msi-vij)/n); vj=max(0,(msj-vij)/m);
if (vj>min(vij,vi)) && (pmiss>0.05) && (LowICC==0) % Test the column effect
    warning('Non-negligible column effect: target ICC possibly biased')
end
iccX = vi/(vi+vij/n);                              % ICC of the data table
iccCor = iccX/(1-pmiss*(1-iccX));       % ICC corrected for missing data
if LowICC>0                                     % Set the target ICC
    iccTar=iccX;
else
    iccTar=iccCor;
end

% ------------------------ Imputation procedure -------------------------
if pmiss>eps

  IX=zeros(m,n);
```

```
if maxmiss<2                              % Deterministic imputation case
  for i=1:m
    x=X(i,:); ix=IX(i,:);
    miss=find(x==MissCode);
    if ~isempty(miss)
      resp=find(x~=MissCode);
      mresp=mean(ix(resp),2);
      ix(miss)=mresp;
    end
  IX(i,:)=ix;
  end                                     % End deterministic imputation
else                                      % Random imputation case
  CIX=IX;
  for j=1:n                                     % ARI on the columns
    x=X(:,j);
    miss=find(x==MissCode);
    if ~isempty(miss)
      resp=find(x~=MissCode);
      r=fix(rand(length(miss),1)*length(resp))+1;
      don=resp(r);
      mdon=mean(x(don));
      mresp=mean(x(resp));
      x(miss)=mresp+(x(don)-mdon);
    end
    CIX(:,j)=x;
  end                                     % End of ARI on the columns
cmin=0; cmax=10; ctol=0.0001;             % Dichotomic search of c
todo=true;
while todo
  c=(cmin+cmax)/2;
  for i=1:m                               % Adjust the rows using current c
    x=X(i,:); ix=CIX(i,:);
    miss=find(x==MissCode);
    if ~isempty(miss)
      resp=find(x~=MissCode);
      mdon=mean(ix(miss),2);
      mresp=mean(ix(resp),2);
      ix(miss)=mresp+c*(ix(miss)-mdon);
    end
    IX(i,:)=ix;
  end                                     % End of adjust the rows
ti = zeros(m,1); ni = ti;                        % ANOVA of IX
tj = zeros(1,n); nj = tj;
sx2 = 0;
for i = 1:m
  for j = 1:n
        ti(i,1) = ti(i,1)+IX(i,j);
        ni(i,1) = ni(i,1)+1;
        tj(1,j) = tj(1,j)+IX(i,j);
        nj(1,j) = nj(1,j)+1;
        sx2 = sx2 + IX(i,j)^2;
  end
end
mitemIX = ti./ni;
N = sum(ni);
t = sum(ti); ss = sx2 - t^2/N;
ssi = sum(ti.^2./ni) - t^2/N;
ssj = sum((tj.^2./nj),2) - t^2/N;
ssij = ss - ssi - ssj;
dfi = m-1; dfj = n-1; dfij = N-1-dfi-dfj;
msi = ssi/dfi;
vij = ssij/dfij; vi = max(0,(msi-vij)/n);
iccIX = vi/(vi+vij/n);                           % Obtained ICC
```

```
    if iccIX>iccTar              % Compare the obtained ICC to the target ICC
      cmin=c;
    else
      cmax=c;
    end
    if (cmax-cmin)<ctol
      todo=false;
    end
  end                                % End of the dichotomic search of c
  end
else
    IX=X;                                      % No missing data case
end
difmit=max(abs(mitemX-mitemIX));              % Verification of item means
if difmit>1e-9
    message=['Item mean inaccuracy: ',num2str(difmit)];
    warning(message)
end
end
```
----------------------------------------------------------------------------------------------------

### Commented example of use of the CRARI program

>> load DLP.mat

*% Load DLP raw data (RTs)*

>> ZDLP=zscore(DLP,0);

*% Transform DLP raw data in Z-scores (the argument "0" is the code for missing data in DLP, but the code for missing data in ZDLP is "inf", since 0 is a possible Z-score value)*

>> [m,n]=size(ZDLP); pmiss=sum(sum(ZDLP==inf))/(m*n)

pmiss = 0.1568

*% This is the proportion of missing data in ZDLP (and in DLP)*

>> [IZDLP,cIZDLP,iccZDLP,iccCorZDLP,iccIZDLP] = CRARI(ZDLP,inf,0);

*% CRARI imputation of the missing data in ZDLP, using ZDLP ICC corrected for missing data (iccCorZDLP) as target ICC (i.e. setting LowICC=0 as third input argument). If one set LowICC=1 as third input argument, then the target ICC is the observed one (iccZDLP).*

>> cIZDLP,iccZDLP,iccCorZDLP,iccIZDLP

cIZDLP = 2.1349          *% Obtained c coefficient*

iccZDLP = 0.8626          *% ICC of the input data table (ZDLP)*

iccCorZDLP = 0.8816          *% ICC corrected for missing data*

iccIZDLP = 0.8816          *% ICC of the output data table (IZDLP)*
----------------------------------------------------------------------------------------------------

**Appendix B.** ICCR2 program (Matlab 7.5 code). Texts at right of "%" are comments.

```
-------------------------------------------------------------------------
function [q,icc,conf,pmiss,iccCor,mitem,r2,r2onICC,r2Cor] = ...
                                ICCR2(X,MissCode,pconf,predictor)
% ---- X ICC, predictor r2, and statistics corrected for missing data -----
%                              Input:
% X:  (m items) x (n participants) data table
% MissCode:  numerical code for missing data in table X (default = inf)
% pconf:  probabilities of ICC confidence intervals (def. [.95 .99 .999])
% predictor: (m items) x (k predictors) table of predictions (optional)
%                              Output:
% q,icc: q ratio and intraclass correlation (ICC) of table X by ANOVA
% conf:  ICC confidence intervals ([prob lower upper])
% pmiss: proportion of missing data in X
% iccCor: ICC corrected for missing data
% mitem: (m x 1) column vector of mean performance for each item
% r2: (1 x k) vector of squared correlations of mitem with the k predictors
% r2onICC: r2/icc (1 x k) vector
% r2Cor: iccCor * r2onICC, squared correlations corrected for missing data
%                              Note:
% Statistics corrected for missing data (*Cor) are reliable only if the
% columns of X have approximately equal means, as with (mixed) Z-scores.
% -------------------------------------------------------------------------

pred=true; if nargin<4, pred=false; end            % Default settings
if nargin<3, pconf=[0.95 0.99 0.999]; end
if nargin<2, MissCode=inf; end

[m,n] = size(X);                                   % ANOVA of X
ti = zeros(m,1); ni = ti; tj = zeros(1,n); nj = tj;
sx2 = 0;
for i = 1:m
    for j = 1:n
        if X(i,j) ~= MissCode
            ti(i,1) = ti(i,1)+X(i,j);
            ni(i,1) = ni(i,1)+1;
            tj(1,j) = tj(1,j)+X(i,j);
            nj(1,j) = nj(1,j)+1;
            sx2 = sx2 + X(i,j)^2;
        end
    end
end
if ~isempty(find(ni==0,1)), error('Error in ICCR2: empty row(s)'), end;
if ~isempty(find(nj==0,1)), error('Error in ICCR2: empty column(s)'), end;
mitem = ti./ni;
N = sum(ni);
pmiss=(m*n-N)/(m*n);                       % Proportion of missing data in X
t = sum(ti); ss = sx2 - t^2/N;
ssi = sum(ti.^2./ni) - t^2/N;
ssj = sum((tj.^2./nj),2) - t^2/N;
ssij = ss - ssi - ssj;
dfi = m-1; dfj = n-1; dfij = N-1-dfi-dfj;
msi = ssi/dfi; msj=ssj/dfj;
vij = ssij/dfij; vi = max(0,(msi-vij)/n); vj=max(0,(msj-vij)/m);
if (vj>min(vij,vi)) && (pmiss>0.05)            % Test of the column effect
    warning('Non-negligible column effect: *Cor statistics not reliable')
end
q = vi/vij; icc = vi/(vi+vij/n); Fobs=msi/vij;         % ICC statistics
Q1f=quantF(1-(1-pconf)/2,dfi,dfij);  % Compute the ICC confidence intervals
Q2f=quantF(1-(1-pconf)/2,dfij,dfi);
```

```
conf=zeros(length(pconf),3);
for i=1:length(pconf)
    conf(i,1)=pconf(i);
    conf(i,2)=1-Q1f(i)/Fobs;
    conf(i,3)=1-1./(Q2f(i)*Fobs);
end
iccCor = icc/(1-pmiss*(1-icc));           % ICC corrected for missing data
if pred                                    % Predictor(s) statitistics
    r2=corrcoef([mitem,predictor]); r2=r2(1,2:end).^2;
    r2onICC=r2/icc;
    r2Cor=iccCor*r2onICC;
else
    r2=[]; r2onICC=[]; r2Cor=[];
end
end

function x = quantF(p,d1,d2)
% F distribution quantiles
x = quantbeta(p,d1/2,d2/2);
x = x.*d2./((1-x).*d1);
end

function x = quantbeta(p,a,b)
% Beta distribution quantiles
tol=1e-6;
x0=zeros(size(p)); x1=ones(size(p));
x=0.5*(x0+x1); dp=betainc(x,a,b)-p;
while max(abs(dp(:)))>tol
    x0(dp<=0)=x(dp<=0); x1(dp>=0)=x(dp>=0);
    x=0.5*(x0+x1); dp=betainc(x,a,b)-p;
end
end
```

--------------------------------------------------------------------------------

Commented example of use of the ICCR2 program

>> [q, icc, conf, pmiss, iccCor, mitem, r2, r2onICC, r2Cor] = ...
               ICCR2(EnglishRT, 0, 0.99, [EnglishLogFreq, EnglishLength]);

*% The "EnglishRT" input argument is a (770 x 94) table of English printed word naming times (raw data from Courrieu et al., 2011) with 3.61% missing data, "0" is the code for missing data, we request a 99% confidence interval for the ICC, and we test two predictors: word log-frequency of use (EnglishLogFreq), and word length in letters (EnglishLength).*

>> q, icc, conf, pmiss, iccCor

q = 0.1333                    *% q ratio of the data table*

icc = 0.9261                  *% ICC of the data table*

conf = 0.9900   0.9160   0.9355   *% 99% confidence interval of the ICC*

pmiss = 0.0361                *% Proportion of missing data in the data table*

iccCor = 0.9286               *% ICC corrected for missing data*

>> r2, r2onICC, r2Cor

r2 = 0.1337   0.0592          *% Squared correlations of the predictors with item mean RTs*

r2onICC = 0.1443   0.0639     *% Squared correlations on ICC ratios*

r2Cor = 0.1340   0.0593       *% Squared correlations corrected for missing data*

-------------------------------------------------------------------------------------------------------

**Acknowledgments**

The authors would like to thank the anonymous reviewers and the action editor, Dr. Ira Bernstein, for their helpful comments. Part of this study was funded by ERC Research Grant 230313.

**Legends and captions**

Table 1: Illustration of the ARI method. In this example, the original dataset for a given item i is composed of five RTs and two missing values (from Participants 3 and 6). During Step 1 of the ARI method, each missing value is replaced by a randomly selected valid data (with replacement). During Step 2, the average of these two replaced RTs (560) is subtracted to each replaced RT. During Step 3, the average of the valid data is added to the previous replaced values. One can verify that the mean RT for the final set of values is equal to the mean RT for the original dataset.

Table 2. ICCs with confidence intervals (95%, 99%, and 99.9%) for four versions of the DLP response time database (Keuleers et al., 2010): the raw data (RTs), the Z-scores, Z-scores with imputation of missing data (16%) by CRARI with the low ICC as target, and Z-scores with imputation of missing data by CRARI with the $\rho_{cor}$ ICC as target. The table also provides the estimate of the $q$ ratio ($q = \sigma_\beta^2 / \sigma_e^2$) corresponding to each ICC. The $c$ coefficients computed by the CRARI algorithm were $c = 2.472$, for the low ICC target, and $c = 2.1349$, for the corrected ICC target.

Figure 1. Plot of the number of items as a function of the number of valid observations (RTs) per item in the DLP database (Keuleers et al., 2010).

Figure 2. ICCs computed from artificial 1400-by-80 Z-score data tables, from an initial complete data table whose ICC was known, and gradually degraded by increasing the proportion of randomly missing data (up to 30%). The "exact" curve corresponds to the ICC of the initial table, the "missing" curve corresponds to the ICCs of the degraded tables, and the "imputed" curve corresponds to the ICCs obtained under imputation of missing data by the ARI method (Chen et al., 2000). The "estimate" curve corresponds to the $\rho_{cor}$ ICCs defined by (7).

Figure 3. Experiments on artificial data similar to those of Figure 2, but before applying Z-score transformations. The "Non-centered columns" plot corresponds to raw data, the "Centered columns" plot corresponds to the same data with zeroed column means, the "Z-scores" plot correspond to full Z-scores, and the "mixed Z-scores" plot corresponds to Z-

scores randomly mixed in each row. The "exact" curve corresponds to the ICC of the initial table, the "missing" curve to the ICCs of degraded tables, and the "estimate" curve to the $\rho_{cor}$ ICCs defined by (7). Note that all plots, except the first one, are identical.

Figure 4. Similar to Figure 2, except that the imputation of missing data was performed by the CRARI method with the $\rho_{cor}$ ICCs defined by (7) as target ICCs.

Figure 5. ECVT test (Courrieu et al., 2011) applied to artificial data that fulfill model (1), with s=0, or that do not fulfill model (1), with s=2. In each case, we tested the original data table, and a degraded version with 16% randomly missing data that were imputed by the CRARI algorithm with the $\rho_{cor}$ ICC defined by (7) as target ICC. Note that the ECVT test provided the correct conclusions in all cases, on imputed data tables as well as on original tables.

Figure 6. ECVT test in an experiment with artificial data as in Figure 5, but using the same table size and the same distribution of missing data as the DLP database (Keuleers et al., 2010). Once again, the ECVT test provided the correct conclusions in all cases, on imputed data tables as well as on original tables.

Figure 7. ECVT test of the DLP database (Keuleers et al., 2010), after Z-score transformation of the data, and imputation of the 16% missing data by the CRARI algorithm with the $\rho_{cor}$ ICC defined by (7) as target ICC. The "predicted" and "observed" curves are undistinguishable, and the $\chi^2$ test is clearly non significant, so the Z-scores of the DLP database fulfill model (1), and the ICC approach is relevant for these data.

Figure 8. Similar to Figure 7, but using the low ICC as target ICC of the CRARI imputation for missing data. The conclusion of the ECVT test is unchanged.

Figure 9. An experiment on artificial data similar to the experiment of Figure 4, but with degraded data tables up to 90% missing data, and the evolution of the correlation between the original item means and the item means of degraded data tables.

Figure 10. Evolution of the $r^2/ICC$ ratio, as a function of the number of participants taken into account, for two predictors (word log-frequency of use, and word length in letters) of word naming times in two languages (English and French), using two empirical databases from Courrieu et al. (2011).

Figure 11. An experiment using artificial data similar to those of Figure 9, and an artificial predictor with know $r^2$ with the original data. The figure plots the observed $r^2$ goodness of fit statistics and the $r^2_{cor}$ estimates as functions of the percentage of missing data in the degraded data tables.

Table 1

| Participant | 1 | 2 | 3 | 4 | 5 | 6 | 7 | means |
|---|---|---|---|---|---|---|---|---|
| Item i data | 500 | 570 | Missing | 630 | 520 | Missing | 620 | 568 |
| Step 1 | 500 | 570 | <u>620</u> | 630 | 520 | <u>500</u> | 620 | <u>560</u> |
| Step 2 | 500 | 570 | 620-<u>560</u> = 60 | 630 | 520 | 500-<u>560</u> = -60 | 620 | |
| Step 3 | 500 | 570 | 60+568 = 628 | 630 | 520 | -60+568 = 508 | 620 | 568 |

Table 2

| DLP data | ICC   (q ratio) | 95% confidence | 99% confidence | 99.9% confidence |
|---|---|---|---|---|
| Raw data (RTs) 16% missing | 0.845 (0.1396) | [0.841, 0.849] | [0.840, 0.850] | [0.839, 0.851] |
| Z-scores 16% missing | 0.863 (0.1610) | [0.859, 0.866] | [0.858, 0.867] | [0.857, 0.868] |
| Z-scores, low ICC imputation | 0.863 (0.1610) | [0.859, 0.866] | [0.858, 0.867] | [0.857, 0.868] |
| Z-scores, corrected ICC imputation | 0.882 (0.1909) | [0.879, 0.884] | [0.878, 0.885] | [0.877, 0.886] |

Figure 1

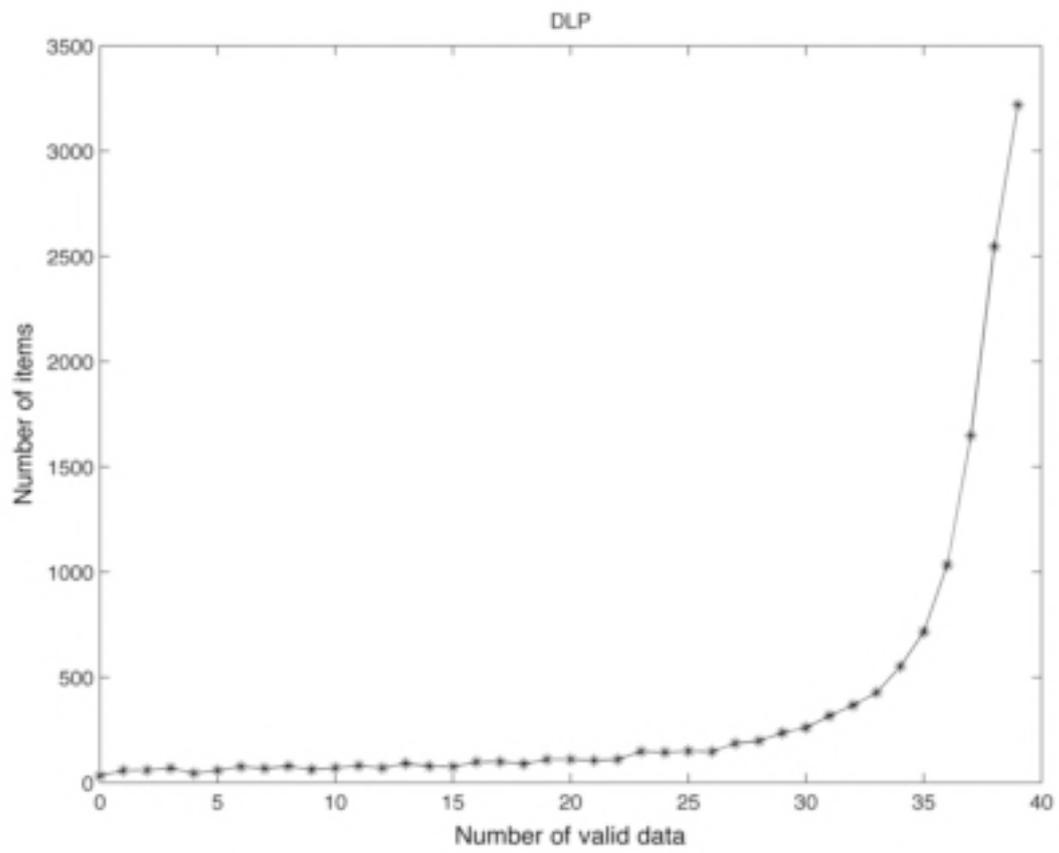

Figure 2

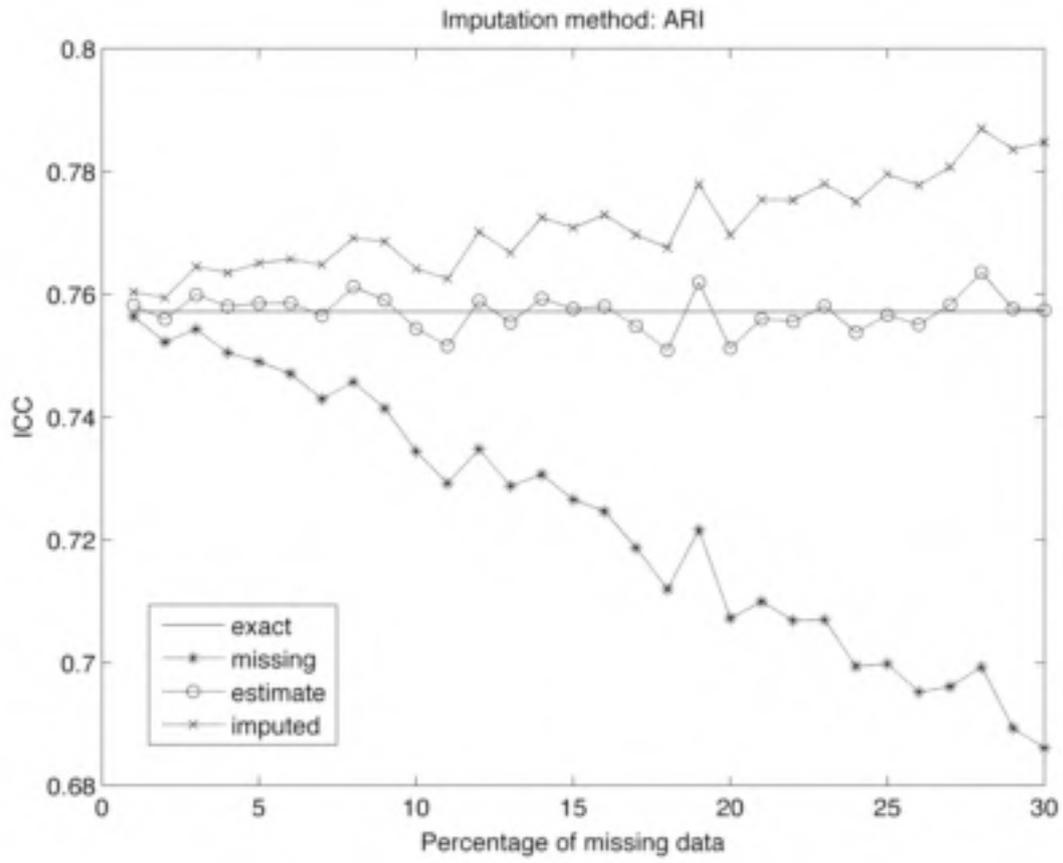

Figure 3

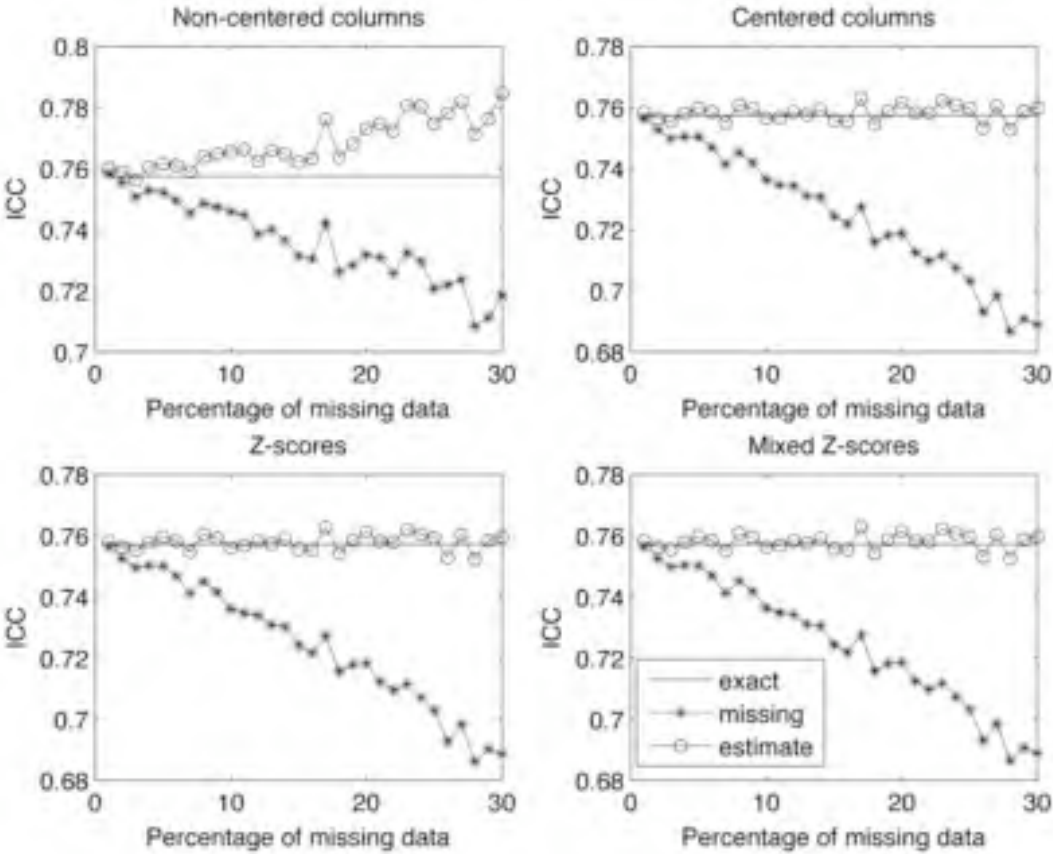

Figure 4

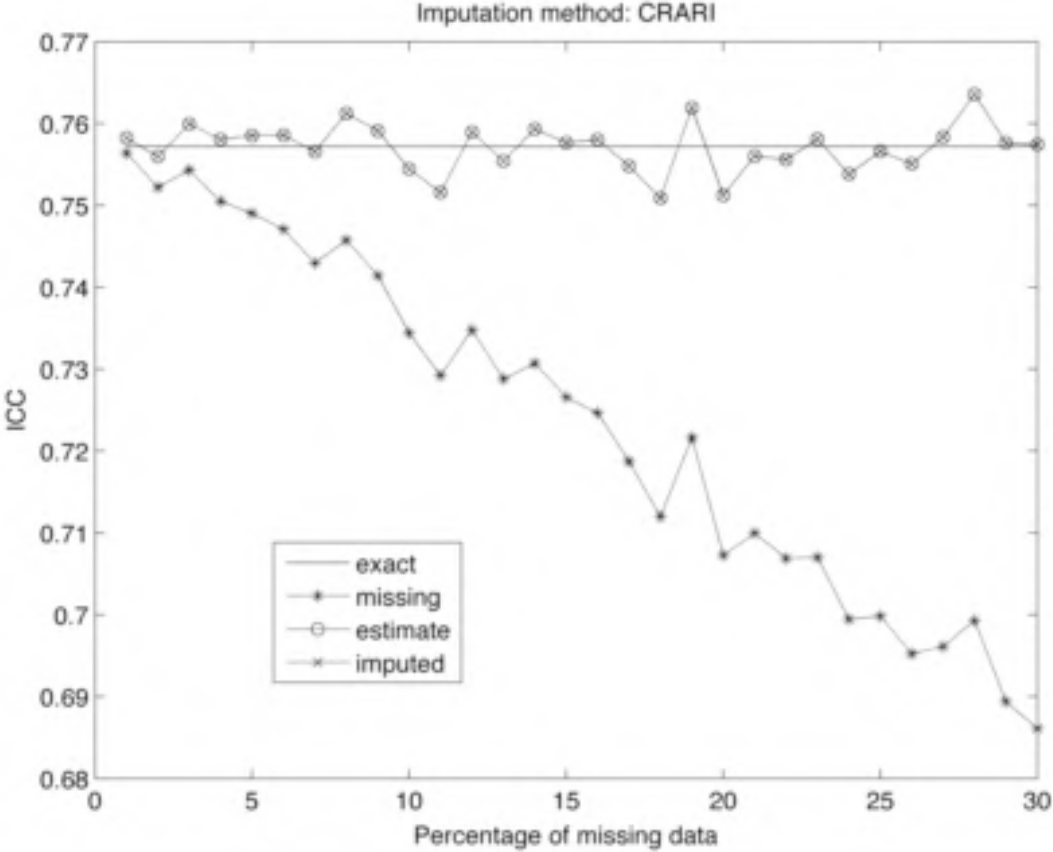

Figure 5

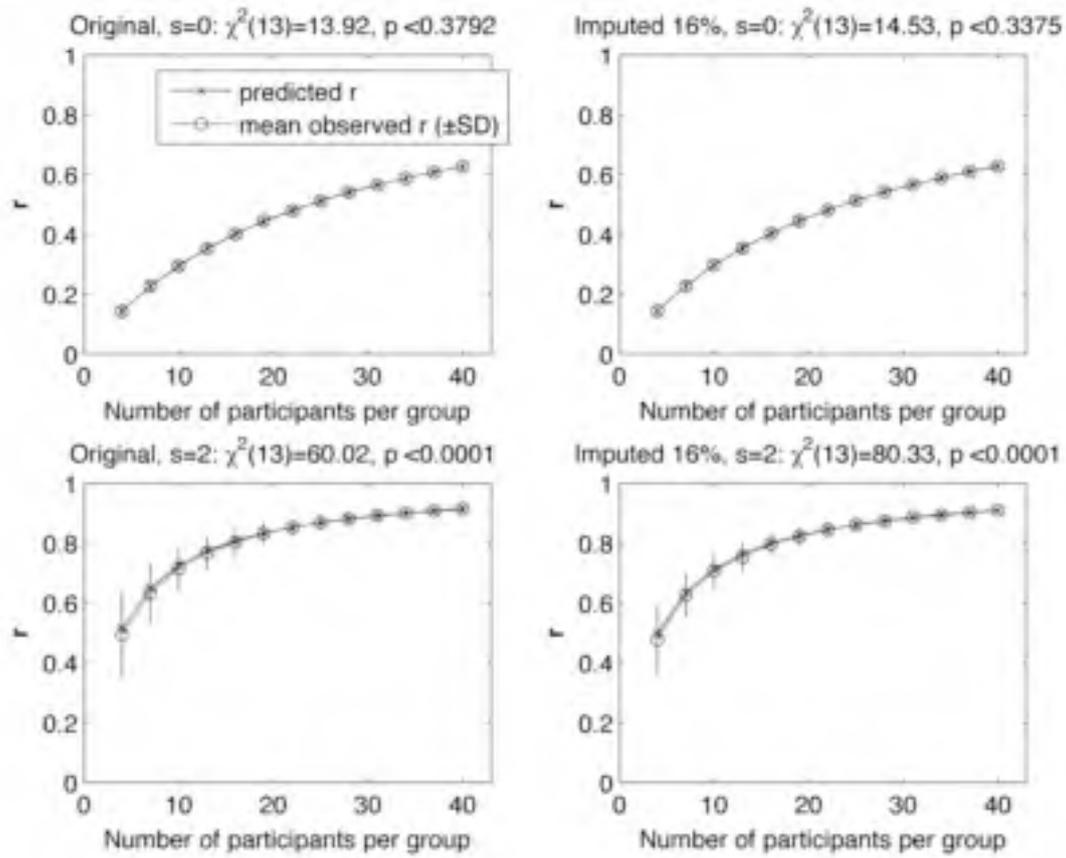

Figure 6

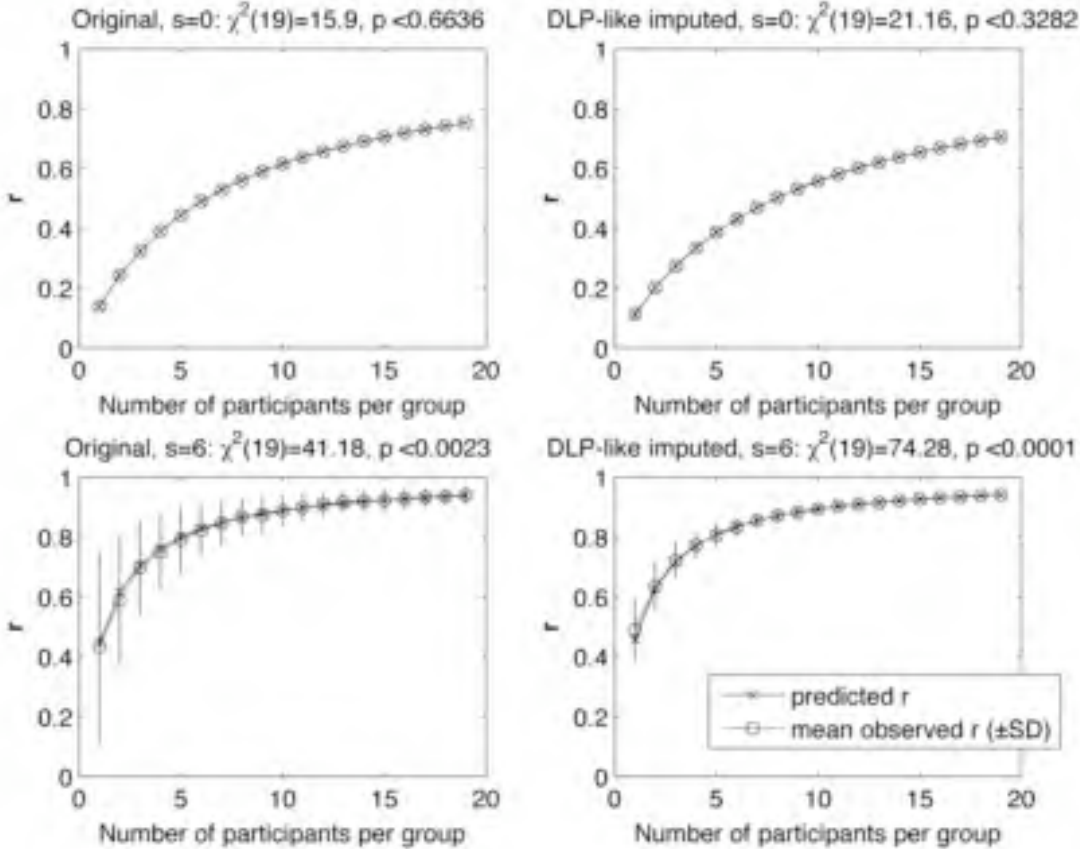

Figure 7

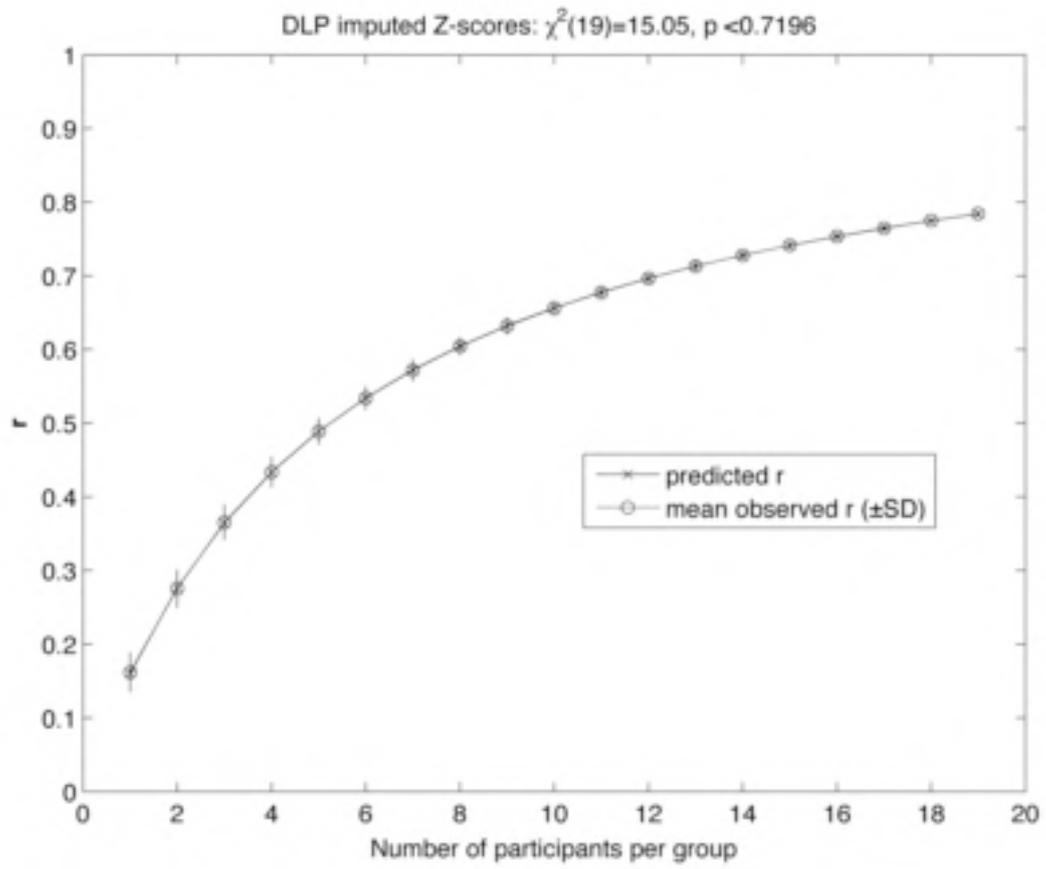

Figure 8

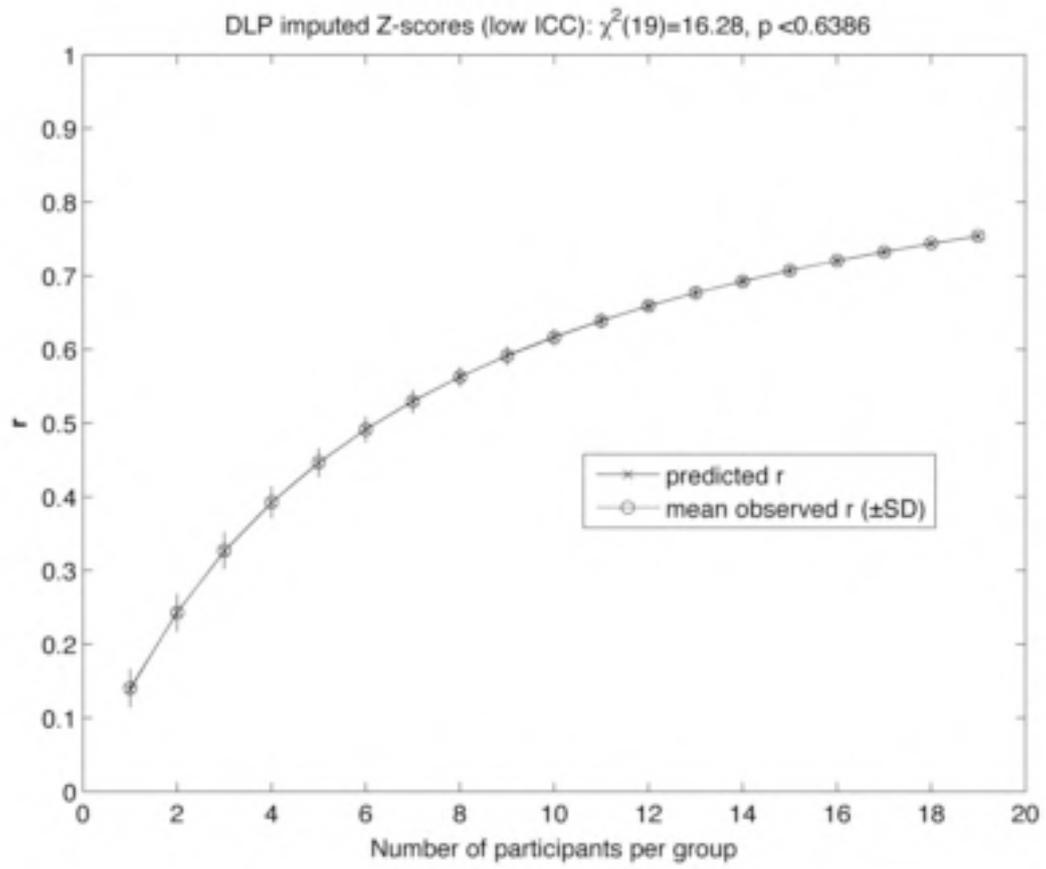

Figure 9

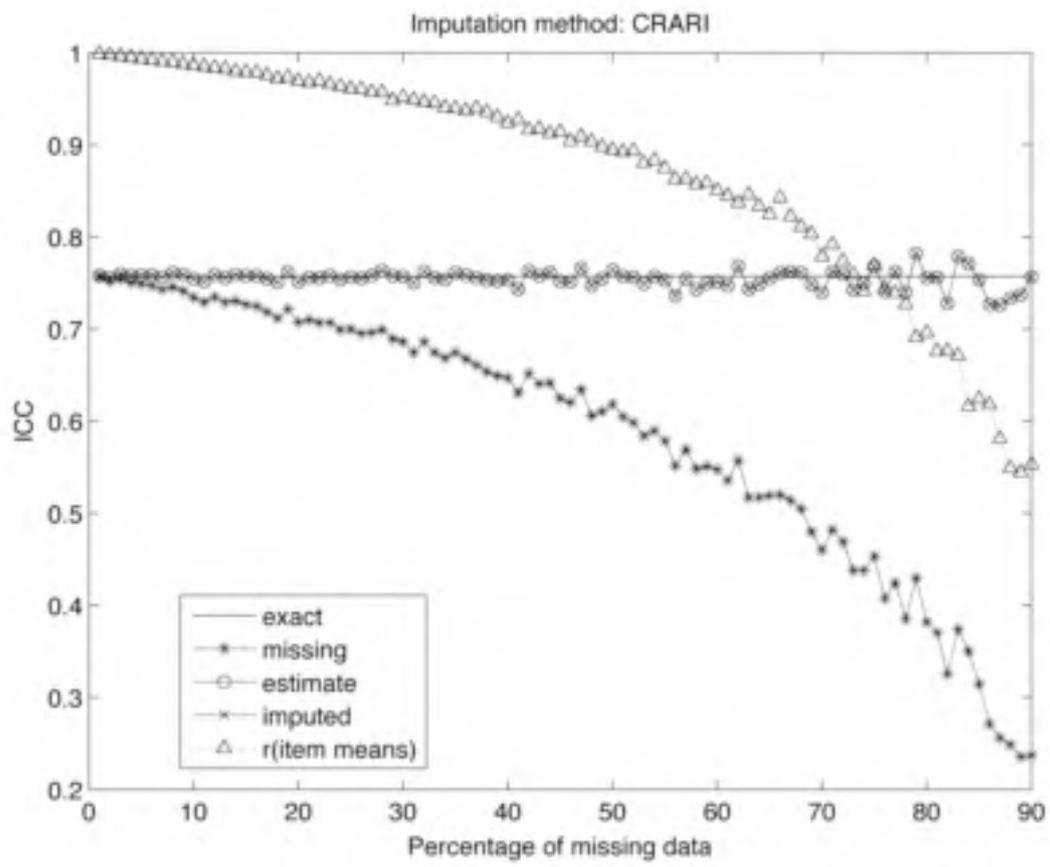

Figure 10

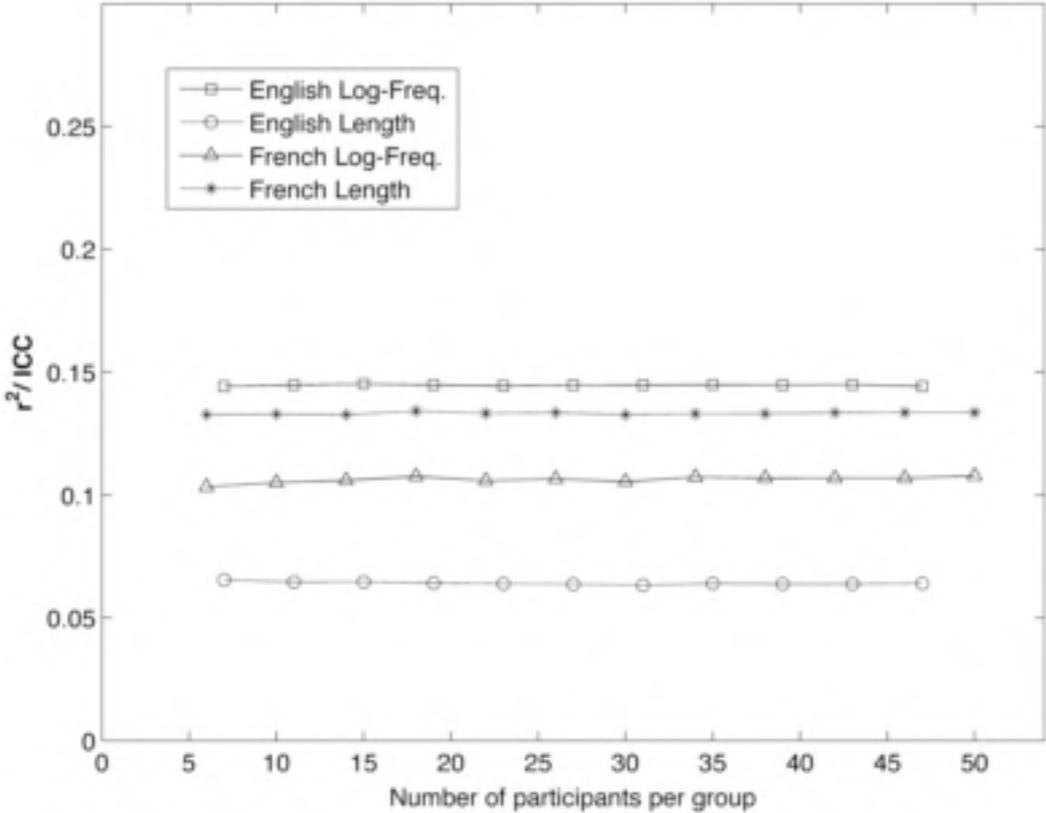

Figure 11

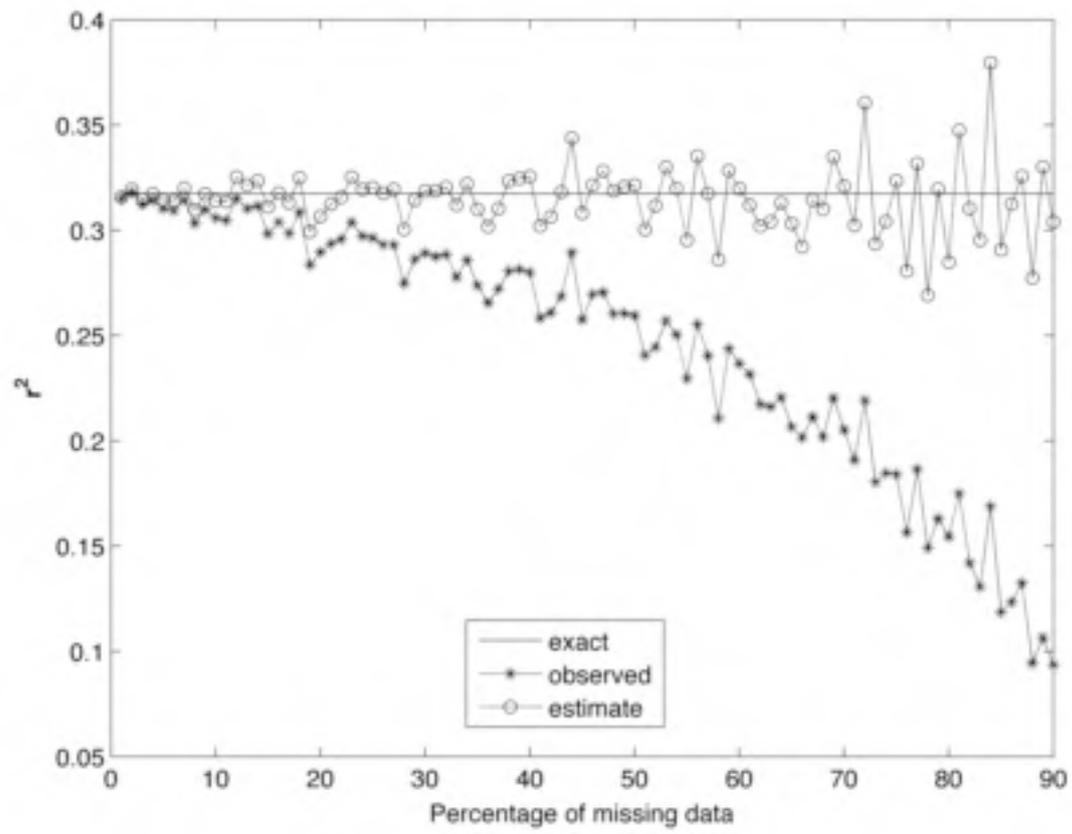